\documentclass{ws-p8-50x6-00}

\begin{document}

\title{Generalized Dipole Polarizabilities of the Pion}

\author{S. Scherer}

\address{Institut f\"{u}r Kernphysik, Johannes Gutenberg-Universit\"{a}t,
J.~J.~Becher-Weg 45, D-55099 Mainz, Germany\\
E-mail: scherer@kph.uni-mainz.de}


\maketitle

\abstracts{We discuss the virtual Compton scattering amplitude
$\gamma^\ast+\pi\to \gamma+\pi$ which enters the reaction 
$\pi^-+e^-\to\pi^-+e^-+\gamma$.
   The low-energy scattering amplitude is divided into a model-independent 
part, involving only the electromagnetic form factor, and a residual part 
which contains structure information related specifically to 
(virtual) Compton scattering.
   In the limit of vanishing final-photon energy, the residual amplitude
can be expressed in terms of three generalized dipole polarizabilities
which are functions of the squared virtual-photon four-momentum.
   We study the VCS amplitude in the framework of chiral perturbation
theory at ${\cal O}(p^4)$.
   At the one-loop level the generalized dipole polarizabilities 
are degenerate $\alpha_L(q^2)=\alpha_T(q^2)=-\beta(q^2)$.}

\section{Introduction}\label{sec:intro}
   Low-energy Compton scattering has a long history of being 
both an important theoretical and experimental testing ground for
models of hadron structure.
   For example, the famous low-energy theorem (LET) of Low \cite{Low_54} 
and Gell-Mann and Goldberger \cite{GellMann_54} 
predicts the scattering amplitude for a spin-$\frac{1}{2}$ system
in terms of the charge, mass, and magnetic moment in the first two 
orders in the photon energy. 
   Terms of second order are no longer determined by the LET and thus 
contain the first information on the structure of the 
nucleon specific to Compton scattering.
   For a general target, these effects can be parametrized in terms of 
two new structure constants, the electric and magnetic polarizabilities
\cite{Klein_55}.

   The LET of real Compton (RCS) scattering has recently been extended to also
include off-shell photons \cite{Scherer_96,Fearing_98}.
   Use of virtual photons substantially increases the possibilities to
investigate the structure of the target, because on the one hand
the energy and three-momentum of the virtual photon can be varied 
independently, and on the other hand also longitudinal components of
current operators become accessible.
   The amplitude for virtual Compton scattering (VCS) off the proton
and the pion can, respectively, be studied in the reactions 
$e^-p\to e^-p\gamma$ and $\pi^-e^-\to\pi^-e^-\gamma$.
   In particular, one is interested in the investigation of generalizations 
of the RCS polarizabilities to the spacelike region, namely, the
so-called generalized polarizabilities 
\cite{Guichon_95,Drechsel_97,Unkmeir_99}.

\section{Formalism and Kinematics}\label{sec:formKin}
\begin{figure}
\begin{center}
\epsfig{file=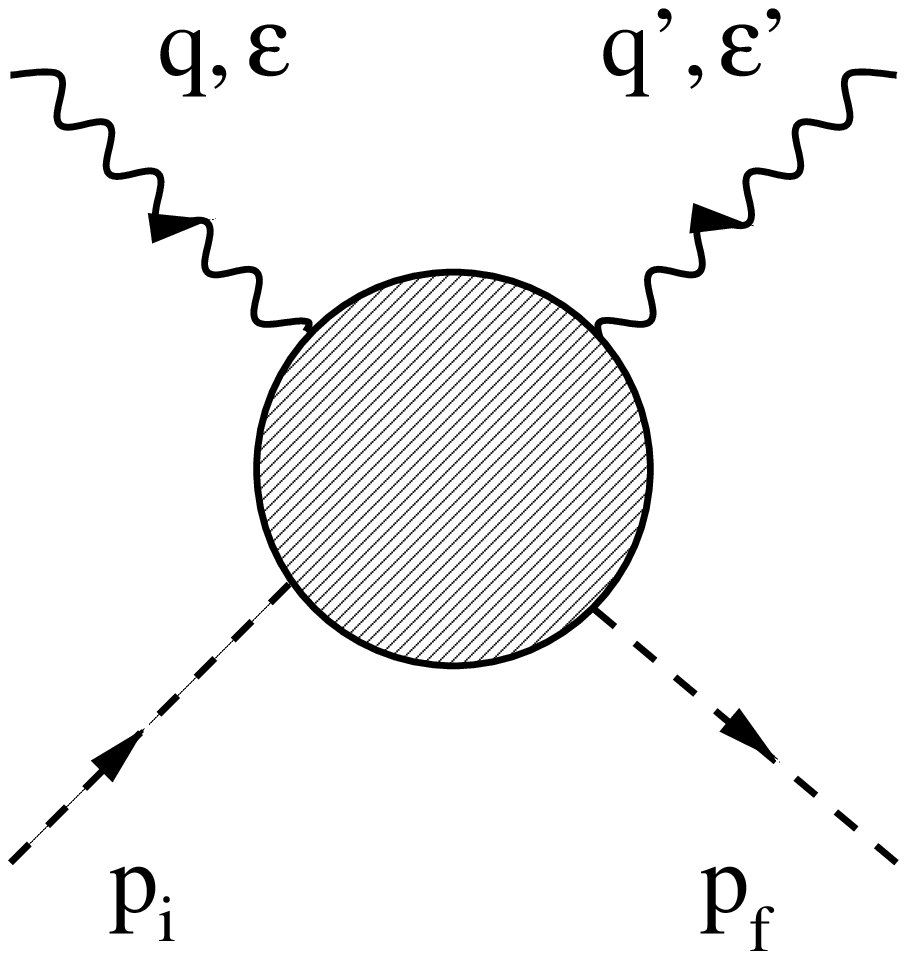,width=4cm}
\caption{\label{comptontensor.fig} Compton scattering off the pion. 
}
\end{center}
\end{figure}
   Let us first define the object of interest, namely, the tensor for Compton 
scattering of off-shell photons (see Fig.\ \ref{comptontensor.fig}):
\begin{eqnarray}
\label{definition:vcstensor}
\lefteqn{(2\pi)^4\delta^4(p_f+q'-p_i-q) M^{\mu\nu}(p_f,q';p_i,q)=}\nonumber\\
&&\int d^4x\, d^4y\, e^{-iq\cdot x}e^{iq'\cdot y}
<\!\pi^\pm(p_f)|T[J^\mu(x)J^\nu(y)]|\pi^\pm(p_i)\!>,
\end{eqnarray}
   where $T$ refers to the covariant time-ordered product. 
   The pions are assumed to be on mass shell.
   As a special case, the RCS amplitude ($q^2=q'^2=0$) is obtained by 
contracting Eq.\ (\ref{definition:vcstensor}) with the polarization vectors 
$\epsilon_\mu$ and $\epsilon'^\ast_\nu$ of the initial and final photons, 
respectively,\footnote{We use $e>0$, $e^2/(4\pi)\approx 1/137$.}
\begin{equation}
\label{mrcs}
{\cal M}=-ie^2\epsilon_\mu\epsilon'^\ast_\nu M^{\mu\nu}(p_f,q';p_i,q).
\end{equation}
   In the laboratory frame ($\vec{p}_i=0$), the low-energy expansion of 
Eq.\ (\ref{mrcs}), using Coulomb gauge, reads
\begin{equation}
\label{mrcsle}
{\cal M}=-2ie^2 \vec{\epsilon}\,'^\ast\cdot\vec{\epsilon}
+2m_\pi i\omega\omega' (4\pi \bar{\alpha} 
\vec{\epsilon}\,'^\ast\cdot\vec{\epsilon}
+4\pi \bar{\beta} \hat{q}\,'\times\vec{\epsilon}\,'^\ast
\cdot\hat{q}\times\vec{\epsilon})
+\cdots.
\end{equation}
   The first term is simply the result for a point particle without internal
structure and the second term parametrizes the lowest-order response
in terms of the electromagnetic polarizabilities $\bar{\alpha}$ and
$\bar{\beta}$.
   
   In the framework of classical electrodynamics, the electric polarizability
denotes the connection between a uniform static electric field and the
induced dipole moment\footnote{Empirical numbers for polarizabilities
are usually given in the Gaussian system. We account for this by
introducing appropriate factors of $4\pi$.}
\begin{equation}
\vec{p}_{ind}=4\pi \alpha \vec{E}.
\end{equation}
   For instance, the polarizability of a dielectric sphere of radius
$R$ and dielectric constant $\epsilon$ is given by $\alpha=
\frac{\epsilon-1}{\epsilon+2} R^3$. \cite{Jackson_75}
\begin{figure}
\begin{center}
\epsfig{file=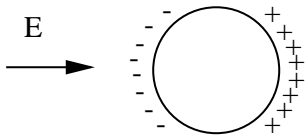,width=4cm}
\caption{\label{elpol:fig} Dielectric sphere in an external electric field.
$\vec{p}_{ind}=4\pi \alpha \vec{E}.$}
\end{center}
\end{figure}
   Using a simple model of a point charge $e$ of mass $m$ bound in
a harmonic oscillator potential, the electric polarizability
$\alpha=e^2/(m\omega^2)$ is easily seen to be 
a measure of the stiffness or rigidity of a system \cite{Holstein_90}.

\begin{figure}
\begin{center}
\epsfig{file=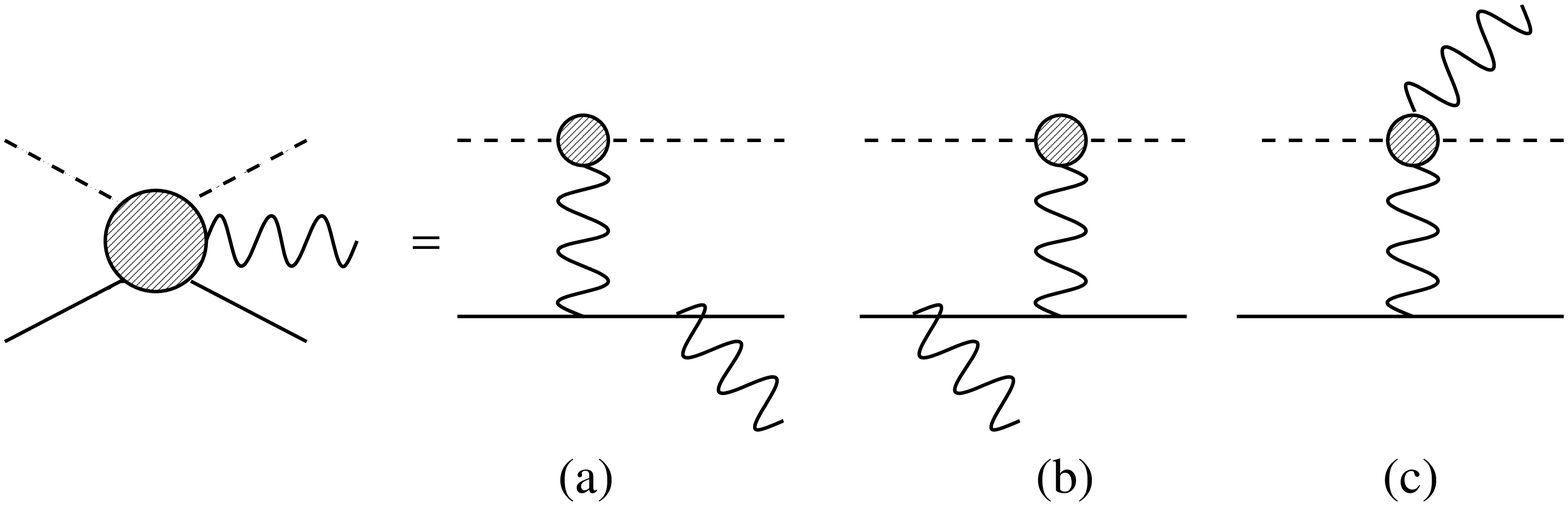,width=10cm}
\caption{\label{diagrams:fig} The reaction 
$\pi^\pm(p_i)+e^-(k)\to \pi^\pm(p_f) + e^-(k')+\gamma(q')$
to lowest order in the electromagnetic coupling: Bethe-Heitler diagrams
(a) and (b), VCS diagram (c).}
\end{center}
\end{figure}

   A more general case of Eq.\ (\ref{definition:vcstensor}), namely
$q^2<0$ and $q'^2=0$, can be studied in the reaction 
$\pi^\pm(p_i)+e^-(k)\to \pi^\pm(p_f) + e^-(k')+\gamma(q')$.
   We will refer to this situation as virtual Compton Scattering (VCS)
because at least one photon is off shell.
   For example, in the Fermilab SELEX E781 experiment inelastic scattering of
high-energy pions off atomic electrons has been measured and 
data are presently analyzed \cite{Moinester_99}.
   At lowest order in the electromagnetic coupling, the relevant diagrams
are shown in Fig.\ \ref{diagrams:fig}. 
   In case of the Bethe-Heitler (BH) diagrams (a) and (b), the real photon is 
emitted by the initial and final electrons.
   It is straightforward to express this contribution in terms of
the electromagnetic form factor of the pion.
   The interesting VCS process is contained as a building block in
diagram (c).

   Since the invariant amplitude is given by the sum of two contributions
${\cal M}_{BH}$ and ${\cal M}_{VCS}$, the differential cross
section is more complicated than in RCS or even in standard electron 
scattering in the one-photon-exchange approximation:
$$d\sigma\sim|{\cal M}_{BH}+{\cal M}_{VCS}|^2.$$
   In particular, the four-momenta exchanged by the virtual photons
in the BH and VCS diagrams read $r\equiv p_f-p_i$ and $q\equiv k-k'=  
r+q'$, respectively.
   Unfortunately, the structure-dependent part of the VCS amplitude is
only a small contribution of the total amplitude.
   However, in principle, the different behavior under the substitution
$\pi^-\to\pi^+$ of ${\cal M}_{\mbox{\footnotesize BH}}$ and 
${\cal M}_{\mbox{\footnotesize VCS}}$, 
\begin{equation}
{\cal M}_{\mbox{\footnotesize BH}}(\pi^-)= 
-{\cal M}_{\mbox{\footnotesize BH}}(\pi^+),\quad
{\cal M}_{\mbox{\footnotesize VCS}}(\pi^-)
= {\cal M}_{\mbox{\footnotesize VCS}}(\pi^+),
\end{equation}
   could be of use in identifying this contribution
by comparing the reactions involving a $\pi^-$ and a $\pi^+$ beam
for the same kinematics:\footnote{This argument works for any particle
which is not its own antiparticle such as the $K^+$ or $K^0$. 
Of course, one could also employ the substitution $e^-\to e^+$.} 
\begin{equation}
d\sigma(\pi^+)-d\sigma(\pi^-)\sim 4 \Re \left[
{\cal M}_{\mbox{\footnotesize BH}}(\pi^+)
{\cal M}^\ast_{\mbox{\footnotesize VCS}}(\pi^+)\right].
\end{equation}

\section{Soft-photon amplitude}
   When studying the low-energy behavior of the VCS amplitude, our first
goal is to isolate, in a gauge-invariant fashion, the model-independent
part from the total amplitude.
   The result can be expressed in terms of gross properties of the 
pion which, in principle, are already known from other experiments.
   To this end, various methods have been devised in the literature which, 
up to separately gauge-invariant analytical terms, all give 
the same result.
   Here, we will shortly summarize the results of Low's method \cite{Low_58}, 
where one starts with the amplitude for radiation from ``external legs.''
   This amplitude contains all the contributions which are non-analytic as 
either $q \rightarrow 0$ or $q'\rightarrow 0$.
   The rationale of this approach is that propagators preceding (following)
the emission or absorption of a photon generate the singularities in the 
soft-photon limit, provided that the particle is on-shell in the final 
(initial) state.
   After a thorough expansion of vertices and propagators and dropping
analytical terms (see Ref.\
\cite{Fearing_98} for details), the result, which we take as 
$M^{\mu\nu}_{1}$, is
\begin{equation}
M^{\mu\nu}_{1} = F(q^2) F(q'^2)\left [
\frac{(2p_i+q)^\mu (2p_f+q')^\nu}{s-m^2_\pi}
+ \frac{(2p_f-q)^\mu (2p_i-q')^\nu}{u-m^2_\pi} \right ],
\end{equation}
where $s=(p_i+q)^2$ and $u=(p_i-q')^2$, and $F$ refers to the electromagnetic
form factor of the pion.
  The next step is to apply the gauge invariance condition to the full 
amplitude
\begin{equation} \label{gauge}
{q}_\mu M^{\mu\nu} = 0,\quad M^{\mu\nu}{q'}_\nu = 0.
\end{equation}
   From $q_\mu M^{\mu\nu}_1 =2 F(q^2)F(q'^2)q^\nu$ and
$M^{\mu\nu}_1 q'_\nu =2 F(q^2)F(q'^2)q'^\mu$ in combination with 
Eq.\ (\ref{gauge}) we obtain as a (minimal) solution $M^{\mu\nu}_2 =
-2 F(q^2)F(q'^2)g^{\mu\nu}$.
   The resulting soft-photon amplitude (SPA) is then
\begin{equation}
\label{spa}
M^{\mu\nu}_{SPA} = F(q^2) F(q'^2)\left [
\frac{(2p_i+q)^\mu (2p_f+q')^\nu}{s-m^2_\pi}
+ \frac{(2p_f-q)^\mu (2p_i-q')^\nu}{u-m_\pi^2} -2g^{\mu\nu} \right ].
\end{equation}
   The total amplitude is given by the soft-photon result of Eq.\ (\ref{spa})
in combination with a separately gauge-invariant analytical residual amplitude.
   By construction $M^{\mu\nu}$, $M^{\mu\nu}_{SPA}$, and the 
residual piece are symmetric with respect to photon 
crossing $q\leftrightarrow-q',\mu\leftrightarrow
\nu$ as well as the substitution $p_i\leftrightarrow -p_f$
(charge conjugation plus particle crossing).
   For $q^2=q'^2=0$, Eq.\ (\ref{spa}) reduces to the result for a
structureless spin-zero particle.

\section{Generalized dipole polarizabilities}
   RCS can be described in terms of two functions depending on two
scalar variables.
   We will now discuss a generalization of the RCS polarizabilities
to the case $q^2\leq 0$ and $q'^2=0$.
   The corresponding invariant amplitude can be parametrized in terms
of three functions $B_1$, $B_2$, $B_5$ which depend on three scalar 
variables \cite{Unkmeir_99}:\footnote{
The most general case $\gamma^\ast\pi\to\gamma^\ast\pi$ requires
five functions depending on four scalar variables.}
\begin{equation}
\label{mvcs2}
-i{\cal M}=B_1 F^{\mu\nu} F'_{\mu\nu} 
+\frac{1}{4}B_2(P_\mu F^{\mu\nu})(P^\rho F'_{\rho\nu})
+\frac{1}{4}B_5(P^\nu q^\mu F_{\mu\nu})(P^\sigma q^\rho F'_{\rho\sigma}). 
\end{equation}
   In Eq.\ (\ref{mvcs2}) $P=p_i+p_f$, and $F^{\mu\nu}$ and $F'_{\mu\nu}$ 
refer to the gauge-invariant combinations
\begin{displaymath}
F^{\mu\nu}=-i q^\mu \epsilon^\nu+iq^\nu \epsilon^\mu,\quad
F'_{\mu\nu}=iq'_\mu\epsilon'^\ast_\nu-iq'_\nu\epsilon'^\ast_\mu.
\end{displaymath}
  Eq.\ (\ref{mvcs2}) becomes particularly simple when
evaluated in the pion Breit frame (p.B.f.) def\/ined by $\vec{P}=0$,
\begin{equation}
\label{mvcsnbf}
-i{\cal M}=\left[2 B_1 \vec{B}\cdot\vec{B}' 
-\left(2B_1+\frac{P^2}{4} B_2\right)
\vec{E}\cdot\vec{E}'
+\frac{P^2}{4}B_5 \vec{q}\cdot \vec{E} \vec{q}\cdot\vec{E}'\right]_{p.B.f.}.
\end{equation}   
   In order to arrive at this equation, we made use
of
\begin{eqnarray*}
F^{\mu\nu} F'_{\mu\nu}&=&[-2\vec{E}\cdot\vec{E}'+2\vec{B}\cdot\vec{B}']_{
p.B.f.},\\
P_\mu F^{\mu\nu} P^\rho F'_{\rho\nu}&=&
[-P^2_0 \vec{E}\cdot\vec{E}']_{p.B.f.},\\
P^\nu q^\mu F_{\mu\nu} P^\rho q^\sigma F'_{\sigma\rho}&=&
[P^2_0 \vec{q}\cdot \vec{E}\vec{q}\cdot\vec{E}']_{p.B.f.},
\end{eqnarray*}   
   where we introduced the notation
$$\vec{E}=i(q_0\vec{\epsilon}-\vec{q}\epsilon_0),\quad
\vec{B}=i\vec{q}\times\vec{\epsilon},\quad
\vec{E}'=-i(q_0'\vec{\epsilon}\,'^\ast-\vec{q}\,'\epsilon_0'^\ast),\quad
\vec{B}'=-i\vec{q}\,'\times\vec{\epsilon}\,'^\ast.
$$
   Note that by def\/inition $[P^2_0]_{p.B.f.}=P^2$.
   Finally, decomposing $\vec{E}=\vec{E}_T+\vec{E}_L$ into components
which are orthogonal and parallel to $\hat{q}$, the parametrization of
the invariant amplitude in the p.B.f. reads 
\begin{eqnarray}
\label{mvcspbf2}
-i{\cal M}&=&\left\{
2 B_1 \vec{B}\cdot\vec{B}' -\left(2B_1+\frac{P^2}{4} B_2\right)
\vec{E}_T\cdot\vec{E}'\right.\nonumber\\
&&\left.
+\left[\frac{P^2}{4}B_5 |\vec{q}|^2-\left(2B_1+\frac{P^2}{4} B_2\right)\right]
\vec{E}_L\cdot\vec{E}'\right\}_{p.B.f.}.
\end{eqnarray} 
   Eq.\ (\ref{mvcspbf2}) serves as the basis of taking the low-energy limit 
$\omega'\to 0$.
   Discussing only the residual amplitudes,\footnote{Given Eq.\ (\ref{spa}),
it is straightforward to evaluate the soft-photon contribution in the
p.B.f.}
 $B_i^r\to b_i^r(q^2)$, it
is natural to def\/ine the following three generalized dipole 
polarizabilities 
\begin{eqnarray}
\label{beta}
8\pi m_\pi\beta(q^2)&\equiv&2b_1^r(q^2),\\
\label{alphat}
8\pi m_\pi
\alpha_T(q^2)&\equiv&-2b_1^r(q^2)-\left(M^2-\frac{q^2}{4}\right)b_2^r(q^2),\\
\label{alphal}
8\pi m_\pi\alpha_L(q^2)&\equiv&-2b_1^r(q^2)-\left(M^2-\frac{q^2}{4}
\right)[b_2^r(q^2)+q^2 b_5^r(q^2)],
\end{eqnarray}
   the superscript $r$ referring to the residual amplitudes beyond the 
soft-photon result.  
   In general, the transverse and longitudinal electric polarizabilities
$\alpha_T$ and $\alpha_L$ will differ by a term, vanishing however in the 
RCS limit $q^2=0$.
   At $q^2=0$, the usual RCS polarizabilities are recovered,
\begin{equation}
\label{rcslimit}
\beta(0)=\bar{\beta},\quad
\alpha_L(0)=\alpha_T(0)=\bar{\alpha}.
\end{equation}
   Observe that $[\vec{B}\cdot\vec{B}']_{p.B.f.}$ and $
[\vec{E}_L\cdot \vec{E}']_{p.B.f.}$ are of ${\cal O}(\omega')$
whereas $[\vec{E}_T\cdot\vec{E}']_{p.B.f.}= {\cal O}(\omega'^2)$,
since $[\vec{E}_T]_{p.B.f}=iq_0(\vec{\epsilon}-\vec{\epsilon}\cdot
\hat{q}\hat{q})={\cal O}(\omega')$.
   In other words, different powers of $\omega'$ have been kept.

\section{Results in chiral perturbation theory}
\subsection{The chiral Lagrangian}
   Chiral perturbation theory \cite{Weinberg_79,Gasser_84} 
provides a natural basis for discussing
low-energy phenomena involving pions, including their interaction
with external fields.
   It is based on a global chiral $\mbox{SU(2)}_L\times \mbox{SU(2)}_R$ 
symmetry of QCD in the limit of vanishing $u$- and $d$-quark masses,
in combination with the assumption of spontaneous symmetry breaking down 
to $\mbox{SU(2)}_V$.
   The chiral symmetry of QCD is mapped onto the most general effective 
Lagrangian in terms of the Goldstone bosons (pions)
\begin{equation}  
  {\cal L}_{\mbox{\footnotesize eff}}
={\cal L}_2+{\cal L}_4+\cdots, 
\end{equation}
where the subscript $2n$ refers to the order in the 
so-called momentum expansion.
    Couplings to external fields, such as the electromagnetic
field, as well as explicit symmetry breaking due to the finite
quark masses, can be systematically taken into account.
   Covariant derivatives and quark-mass terms count as ${\cal O}(p)$ 
and ${\cal O}(p^2)$, respectively.
   Weinberg's power counting scheme allows for a classification
of Feynman diagrams by establishing a relation between the momentum
expansion and the loop expansion.

    The most general chiral Lagrangian at ${\cal O}(p^2)$ is given by
\begin{equation}
\label{l2}
{\cal L}_2 = \frac{F^2}{4} \mbox{Tr} \left[ D_{\mu} U (D^{\mu}U)^{\dagger} 
+\chi U^{\dagger}+ U \chi^{\dagger} \right],
\end{equation}
where $U$ is a unimodular unitary $(2\times 2)$ matrix containing the
pion fields.
   As a parametrization of $U$ we use   
\begin{equation}
\label{paru}
U(x)=\frac{\sigma(x)+i\vec{\tau}\cdot\vec{\pi}(x)}{F}, 
\quad \sigma^2(x)+\vec{\pi}^2(x)=F^2,
\end{equation}
where $F$ denotes the pion-decay constant in the chiral limit: 
$F_\pi=F[1+{\cal O}(\hat{m})]=92.4$ MeV. 
   In the isospin-symmetric limit $m_u=m_d=\hat{m}$, the 
quark mass is contained in $\chi=2 B_0 \hat{m}=m^2_\pi$
at ${\cal O}(p^2)$, where $B_0$ is related to the quark condensate
$<\!\!\bar{q}q\!\!>$.
   The covariant derivative $D_\mu U = \partial_\mu U +\frac{i}{2}e A_\mu 
[\tau_3,U]$ contains the coupling to the electromagnetic f\/ield $A_\mu$.
   The most general structure of ${\cal L}_4$ was f\/irst obtained by 
Gasser and Leutwyler (see Eq.\ (5.5) of Ref.\ \cite{Gasser_84}) 
and contains seven new low-energy constants $l_i$,
\begin{eqnarray}
\label{l4gl}
{\cal L}_4&=&\cdots +l_5\left[\mbox{Tr}(F^R_{\mu\nu} U F_L^{\mu\nu} U^\dagger)
-\frac{1}{2}\mbox{Tr}(F_{\mu\nu}^L F_L^{\mu\nu}
+F_{\mu\nu}^R F_R^{\mu\nu})
\right]\nonumber
\\
&&+i\frac{l_6}{2}\mbox{Tr}\left(
F^R_{\mu\nu}D^\mu U(D^\nu U)^\dagger
+F^L_{\mu\nu}(D^\mu U)^\dagger D^\nu U\right)+\cdots,
\end{eqnarray}
where $F^R_{\mu\nu}=F^L_{\mu\nu}=-\frac{e}{2}\tau_3
(\partial_\mu A_\nu-\partial_\nu A_\mu)$.

\subsection{The soft-photon amplitude}
   Our goal is to evaluate the VCS amplitude at the one-loop level which
corresponds to ${\cal O}(p^4)$ in the momentum expansion.
   At ${\cal O}(p^4)$, the result for the s- and u-channel pole terms
reads
\begin{eqnarray}
\label{mp}
{\cal M}_P&=&-ie^2\left\{F(q^2)\left[\frac{2p_f\cdot\epsilon'^\ast\,
(2p_i+q)\cdot\epsilon}{s-m_\pi^2}
+\frac{(2p_f-q)\cdot\epsilon\, 2p_i\cdot\epsilon'^\ast}{u-m_\pi^2}\right]
\right.\nonumber\\
&&\left.
+2 q\cdot\epsilon\, q\cdot\epsilon'^\ast \frac{1-F(q^2)}{q^2}\right\},
\end{eqnarray}
   where $F(q^2)$ is the prediction for the pion electromagnetic form factor.
   In order to obtain Eq.\ (\ref{mp}), we made use of the electromagnetic
vertex at ${\cal O}(p^4)$
\begin{equation}
\label{gammamu}
\Gamma^\mu(p',p)=(p'+p)^\mu F(q^2)+(p'-p)^\mu \frac{p'^2-p^2}{q^2}[
1-F(q^2)],\quad q=p'-p.
\end{equation} 
   The second part of Eq.\ (\ref{mp}) is analytical as $q\to 0$ 
and would have been dropped in the first step of Low's method.
   It originates from the term proportional to $q^\mu$ in 
Eq.\ (\ref{gammamu}). 
   Observe that Eq.\ (\ref{mp}) is not gauge invariant.
   
\begin{figure}
\begin{center}
\epsfig{file=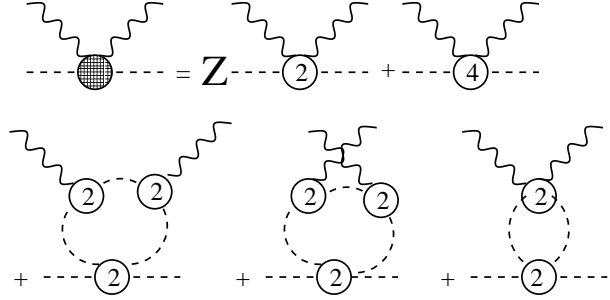,width=8cm}
\caption{\label{classb.fig} One-particle-irreducible diagrams. The labels
$2n$ in the interaction vertices denote the order of the corresponding
Lagrangian. $Z$ is the wave function renormalization constant.}
\end{center}
\end{figure}

   The one-particle-irreducible diagrams of Fig.\ \ref{classb.fig} give rise 
to a residual part of the form
\begin{equation}
\label{mr}
{\cal M}_R=ie^2\left[2\epsilon\cdot\epsilon'^\ast
+2(q^2\epsilon\cdot\epsilon'^\ast-q\cdot \epsilon\,
q\cdot\epsilon'^\ast)\frac{F(q^2)-1}{q^2}\right]+\tilde{\cal M}_R.
\end{equation}
   Combining Eqs.\ (\ref{mp}) and (\ref{mr}), 
${\cal M}={\cal M}_P+{\cal M}_R=\tilde{\cal M}_P+\tilde{\cal M}_R$, 
one finds that $\tilde{\cal M}_P$ is exactly of the form predicted 
by Eq.\ (\ref{spa}) for $q'^2=0$. 
   In particular, $\tilde{\cal M}_P$ and $\tilde{\cal M}_R$ 
are now separately gauge invariant.

\subsection{The residual amplitude}
\begin{figure}
\begin{center}
\epsfig{file=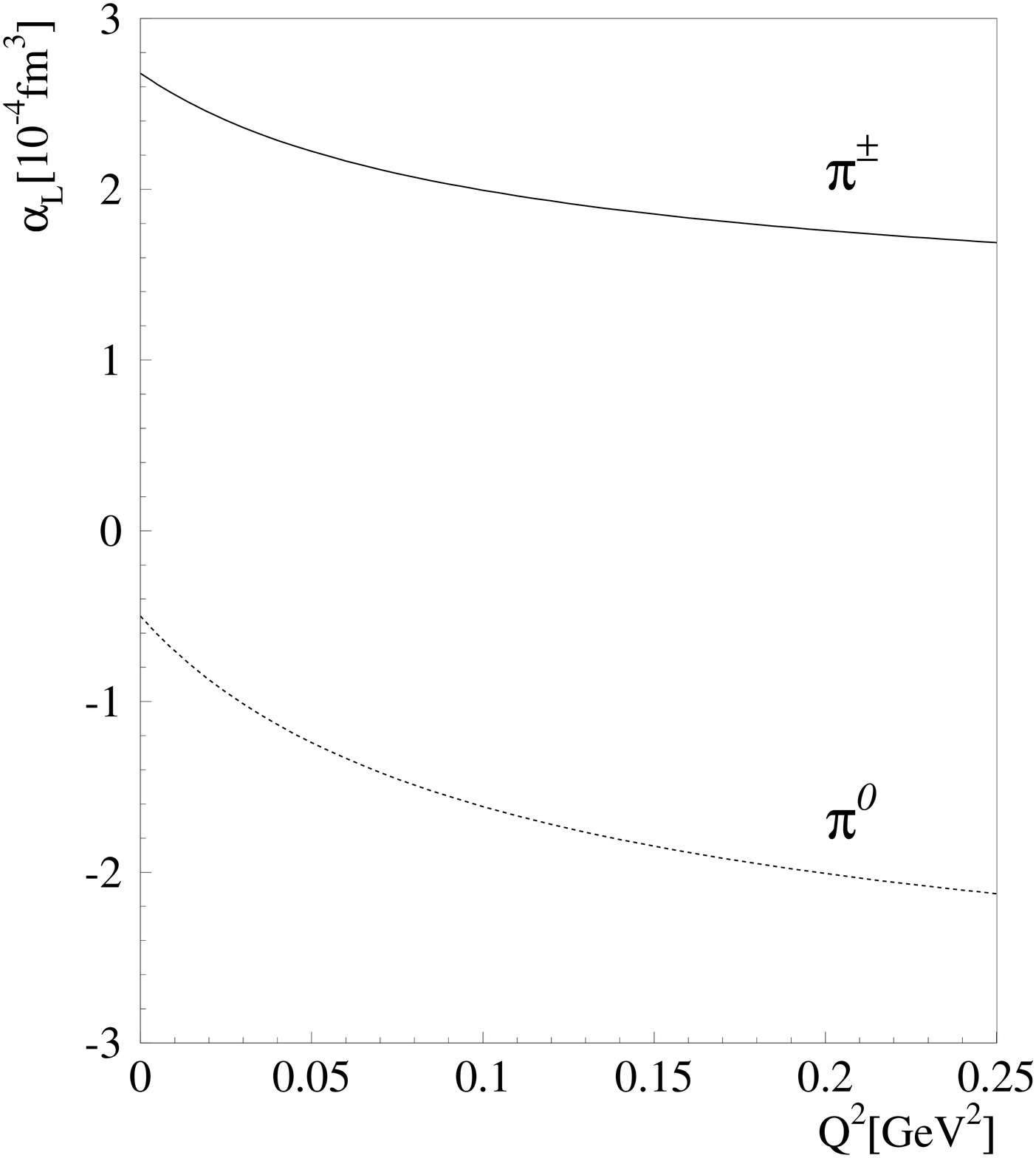,width=8cm}
\caption{\label{alpha.fig}
${\cal O}(p^4)$ prediction for the generalized dipole polarizabilities
$\alpha_L(-Q^2)$ of the charged pion (solid curve) and the neutral pion
(dashed curve) as function of $Q^2$.
   At ${\cal O}(p^4)$,  $\alpha_L(q^2)=\alpha_T(q^2)=-
\beta(q^2)$.}
\end{center}
\end{figure}

   At ${\cal O}(p^4)$, the result for the residual amplitude 
$\tilde{\cal M}_R$ reads
\begin{equation}
\label{mrtilde}
\tilde{\cal M}_R=-ie^2(q'\cdot\epsilon \,q\cdot \epsilon'^\ast-q\cdot q'
\epsilon\cdot\epsilon'^\ast)
\left[-\frac{4(2 l_5^r-l_6^r)}{F^2_\pi}
+\frac{2q\cdot q'-q^2}{16\pi^2 F^2_\pi q\cdot q'}
{\cal G}(q^2,q\cdot q')\right],
\end{equation}
   where ${\cal G}(q^2,q\cdot q')$ is a one-loop function given in
Eq.\ (26) of Ref.\ \cite{Unkmeir_99}.
   The combination $2l^r_5-l^r_6=(2.85\pm 0.42)\times 10^{-3}$ is
determined  through the decay $\pi^+\to e^+\nu_e\gamma$.
   Using the results of Sec.\ 4 it is straightforward to extract
the generalized dipole polarizabilities
\begin{eqnarray}
\label{alphapp2}
\alpha^{\pi^\pm}_L(q^2)&=&
\alpha^{\pi^\pm}_T(q^2)=
-\beta^{\pi^\pm}(q^2)\nonumber\\
&=&
\frac{e^2}{8\pi m_\pi}
\left[\frac{4(2l^r_5-l^r_6)}{F^2_\pi}-\frac{q^2}{m_\pi^2}
\frac{1}{(4\pi F_\pi)^2} J^{(0)'}\left(\frac{q^2}{m_\pi^2}\right)\right],
\end{eqnarray}
   where 
$${J^{(0)}}'(x)=\frac{1}{x}\left[1-\frac{2}{x\sigma(x)}\ln\left(
\frac{\sigma(x)-1}{\sigma(x)+1}\right)\right],
\quad \sigma(x)=\sqrt{1-\frac{4}{x}},
\quad x\le 0.
$$
     The results for the generalized dipole polarizabilities are shown
in F\/ig.\ \ref{alpha.fig}.\footnote{For the sake of completeness, 
neutral pion polarizabilities are also
shown \cite{Unkmeir_99}.}
   The $q^2$ dependence does not contain any ${\cal O}(p^4)$ parameter,
i.e., it is entirely given in terms of the pion mass and the pion decay 
constant $F_\pi=92.4$ MeV.
   At $q^2=0$, the electromagnetic polarizabilities of the charged pion 
are determined by
an ${\cal O}(p^4)$ counter term \cite{Holstein_90,Terentev_73}, 
\begin{equation}
\label{t:ss:alpha}
\bar{\alpha}=-\bar{\beta}=\frac{e^2}{4\pi} \frac{2}{m_\pi F^2_\pi}
(2l^r_5-l^r_6)
=(2.68\pm0.42)\times 10^{-4}\,\mbox{fm}^3.
\end{equation}
   Corrections to this result at ${\cal O}(p^6)$ were shown
to be reasonably small, namely 12\% and 24\% of the ${\cal O}(p^4)$ values
for $\bar{\alpha}$ and $\bar{\beta}$, respectively \cite{Buergi_96}.
   Empirical results for the RCS polarizabilities have been obtained
from high-energy pion-nucleus bremsstrahlung, 
$\bar{\alpha}_{\pi\pm}=(6.8\pm 1.4)\times 10^{-4}\,\mbox{fm}^3$ 
\cite{Antipov_83},
$\bar{\beta}_{\pi^\pm}=(-7.1\pm 4.6)\times 10^{-4}\,\mbox{fm}^3$
\cite{Antipov_85},
and radiative pion photoproduction off the nucleon,
$\bar{\alpha}_{\pi\pm}=(20\pm 12)\times 10^{-4}\,\mbox{fm}^3$ 
\cite{Aibergenov_86}.
   An improved accuracy is required to test the chiral predictions.

   As in the case of RCS, we expect the degeneracy
$\alpha_L(q^2)=\alpha_T(q^2)=-\beta(q^2)$ to be lifted at the
two-loop level.

\section{Summary}
   
   We discussed the invariant amplitude for virtual Compton scattering
off the charged pion $\gamma^\ast+\pi\to \gamma +\pi$ which enters the reaction
$\pi+e\to \pi +e+\gamma$.
   In the framework of Low's method we derived the soft-photon result
which depends only on the electromagnetic form factor of the pion.
   From the analysis of a covariant approach, evaluated in the pion Breit 
frame, we introduced three generalized dipole polarizabilities
which are functions of the squared virtual-photon momentum transfer.
   
   We then discussed VCS of the pion in the framework of ChPT at 
${\cal O}(p^4)$.
   As expected, the ChPT result reproduces the soft-photon amplitude.
   The momentum dependence of the generalized polarizabilities is entirely 
predicted in terms of the pion mass and the pion-decay constant, i.e., 
no additional counter-term contribution appears.
   The predictions at ${\cal O}(p^4)$ show a degeneracy of the 
polarizabilities, $\alpha_L(q^2)=\alpha_T(q^2)=-\beta(q^2)$.
   As in the case of real Compton scattering, we expect the degeneracy
to be removed at the two-loop level.

\section*{Acknowledgments}
   This work was supported by the Deutsche Forschungsgemeinschaft (SFB 443).
   The author would like to thank D.\ Drechsel, H.W.\ Fearing, 
A.I.\ L'vov, B.\ Pasquini, and C.\ Unkmeir for a pleasant and fruitful 
collaboration on various topics related to virtual Compton scattering. 
   It is pleasure to thank M.A.\ Moinester and A.\ Ocherashvili for useful 
discussions on experimental issues in VCS.


\begin{thebibliography}{99}

\bibitem{Low_54} F.E. Low, \Journal{\em Phys. Rev.}{96}{1428}{1954}.

\bibitem{GellMann_54} M. Gell-Mann and M.L. Goldberger, 
\Journal{\em Phys. Rev.}{96}{1433}{1954}.
 
\bibitem{Klein_55} A. Klein, \Journal{\em Phys. Rev.}{99}{998}{1955}.

\bibitem{Scherer_96} S. Scherer, A.Yu. Korchin, and J.H. Koch,
\Journal{\em Phys. Rev. C}{54}{904}{1996}.

\bibitem{Fearing_98} H.W. Fearing and S. Scherer, 
\Journal{\em Few-Body Syst.}{23}{111}{1998}.

\bibitem{Guichon_95} P.A.M. Guichon, G.Q. Liu, and A.W. Thomas,
\Journal{\em Nucl. Phys.}{A591}{606}{1995}. 

\bibitem{Drechsel_97}D. Drechsel, G. Kn\"{o}chlein, A. Metz, and S. Scherer,
\Journal{\em Phys. Rev. C}{55}{424}{1997}.


\bibitem{Unkmeir_99} C. Unkmeir, S. Scherer, A.I. L'vov, and D. Drechsel,
{\tt hep-ph/9904442}.

\bibitem{Jackson_75} J.D. Jackson, {\em Classical Electrodynamics}
(John Wiley, New York, 1975) Sect. 4.4.
 
\bibitem{Holstein_90} B.R. Holstein, 
\Journal{\em Comments Nucl. Part. Phys.}{19}{221}{1990}.

\bibitem{Moinester_99} M.A. Moinester {\em et al.} (The SELEX Collaboration), 
{\tt hep-ex/9903039}.

\bibitem{Low_58} F.E. Low, \Journal{\em Phys. Rev.}{110}{974}{1958}.

\bibitem{Weinberg_79} S. Weinberg, \Journal{\em Physica}{96A}{327}{1979}.

\bibitem{Gasser_84} J. Gasser and H. Leutwyler, 
\Journal{\em Ann. Phys. (N.Y.)}{158}{142}{1984}.

\bibitem{Terentev_73} M. V. Terent'ev,
\Journal{\em Sov. J. Nucl. Phys.}{16}{87}{1973}.

\bibitem{Buergi_96}U. B\"{u}rgi, 
\Journal{\em Phys. Lett. B}{377}{147}{1996};
\Journal{\em Nucl. Phys.}{B479}{392}{1996}. 

\bibitem{Antipov_83} Yu.M. Antipov {\em et al.}, 
\Journal{\em Phys. Lett.}{121B}{445}{1983}.

\bibitem{Antipov_85}Yu.M. Antipov {\em et al.},
\Journal{\em Z. Phys. C}{26}{495}{1985}.

\bibitem{Aibergenov_86} T.A. Aibergenov {\em et al.}, 
\Journal{Czech. J. Phys.}{B36}{948}{1986}.


\end{thebibliography}
\end{document}